\documentclass[12pt]{article}
\usepackage{epsfig}
\usepackage{a4,isolatin1}
\begin{document}
\begin{center}
{\bf \large Causality, particle localization and positivity of the
  energy\footnote{To appear in: {\em Irreversibility and Causality in
Quantum Theory -- Semigroups and Rigged Hilbert Spaces}, edited by 
A. Bohm, H.-D. Doebner and P. Kielanowski, Springer Lecture Notes in
    Physics, Vol. 504 (1998). } \normalsize}
\end{center}

\vspace*{1cm}

\begin{center}
{\bf Gerhard C. Hegerfeldt}\\
Institut f\"ur Theoretische Physik\\
Universit\"at G\"ottingen\\
Bunsenstr. 9\\
D-37073 G\"ottingen
\end{center}

\vspace*{1cm}

\begin{center}
{\bf Abstract}
\end{center}
Positivity of the Hamiltonian alone is used to show that particles, if
initially localized in a finite region, immediately develop infinite
tails.

%\pagebreak
\vspace*{1.3cm}

\noindent {\bf 1. Introduction}

\vspace*{.7cm}

The concept of finite signal velocity or, more precisely, the speed of
light as highest signal velocity is often called Einstein
causality. If, in the special theory of relativity, there were
superluminal signals then there could exist tachyons, i.e. superluminal
particles, and the sequence of cause and effect could be reversed. On
the other hand, if signals of arbitrarily high
velocities existed, one could also argue that one could obtain 
absolute clock synchronization and absolute simultaneity, thus making
a revision of special relativity necessary. 

In usual quantum mechanics it is well known that wave functions, if
initially localized in a finite region, immediately develop infinite
tails. For a nonrelativistic theory this is of no concern. A similar
phenomenon was then noted, however, for the Newton-Wigner position
operator \cite{I1,II6,Ru}. In fact, the question of localization in
quantum field was recognized as a difficult problem quite early
\cite{II1}. In particular, localization by means of a current-density
four-vector were investigated in a class of models \cite{II5}.

It was then recognized by the present author that the superluminal
spreading had nothing to do with position operators, field equations,
current densities or a particular notion of particle localization. In
a model-independent way the theorem was proved that a free
relativistic particle, if initially localized with probability 1 in a
finite (bounded) region, immediately thereafter would have spread over all
space \cite{II7}. An alternative proof of this theorem was given in
\cite{II8} and a generalization to relativistic systems in
\cite{II9}. The theorem was carried over to quite general
interactions, not necessarily relativistic ones in \cite{II10}. It became
apparent in that paper that the main ingredient was positivity of the
energy, with translation invariance used as a technical tool.

In 1985 the present author \cite{II11} showed that the connection
between localization and Einstein causality was more restrictive than
previously thought. It was proved in \cite{II11} that relativistic
systems with Gaussian-like bounded tails at $t =0$ -- provided they
exist! -- would lead to a superluminal probability flow.

The main purpose of this paper is to show that translation invariance
is not needed for superluminal spreading of particles which are initially
confined in a bounded region. For this purpose a recent result \cite{He65}
of the present author for Fermi's two-atom system will be reformulated
in Section 2 as an abstract mathematical theorem and applied to
particle localization. It will then be shown that, as a consequence of
positivity of the energy alone, a particle which is initially strictly
localized in a finite region either stays there indefinitely or
immediately develops infinite tails. In the last section the
connection to Einstein causality is discussed. As in the author's
result \cite{He65} on the Fermi problem several ways out are mentioned
to avoid a conflict.

\vspace*{1.3cm}

\noindent {\bf 2. A consequence of positivity of the energy}

\vspace*{0.7cm}

In this section we will prove a simple mathematical result on the
temporal behavior of certain expectation values. If the
time-development operator is positive one might expect analyticity
properties, but for arbitrary expectation values this is not true. One
has, however, the following result.\\[0.5cm]
{\bf Theorem:} Let $H$ be a selfadjoint operator, positive
or bounded from below, in a Hilbert space ${\cal H}$. For given $\psi_0
\in {\cal H}$ let $\psi_t, t \in I\!\!R$, be defined as 
\begin{equation}\label{t1}
\psi_t = e^{-i Ht} \psi_0~.
\end{equation}
Let $A$ be a positive operator in ${\cal H}, A \ge 0$, and let $p_A(t)$
be defined as
\begin{equation}\label{t2}
p_A(t) = \langle \psi_t , A \psi_t \rangle~.
\end{equation}
Then either
\begin{equation}\label{t3}
p_A(t) \neq 0~~~\mbox{for almost all}~~~ t
\end{equation}
and the set of such $t$'s is dense and open, or
\begin{equation}\label{t4}
p_A(t) \equiv 0~~~\mbox{for all}~~~t~.
\end{equation}

\vspace*{.5cm}

The proof is based on an analyticity argument for which, however, a
little care -- and the positivity of $A$ -- is needed. Evidently,
since $H \geq - c$, one can define $\exp \{- i H (t + i
y)\}~\mbox{for}~ y \le 0$, and $\exp \{- i H z\}$ is analytic in $z$
for Im $z < 0$, and hence $\psi_t$ can be analytically continued to
the lower half-plane, with continuous boundary values on the real
axis. However, the r.h.s. of Eq. (\ref{t2}) can in general not be
analytically continued since it equals
\[
\langle \psi, e^{i H t} A e^{- i H t} \psi \rangle
\]
and since $\exp \{i (H + i y) t\}$ is in general unbounded for $y <
0$. To by-pass this the positivity of $A$ can be used. We write
\begin{equation}\label{t5}
p_A(t) = \langle A^{1/2} \psi_t , A^{1/2} \psi_t \rangle
\end{equation}
where $A^{1/2}$ is the positive square root of $A$, and denote by
${\cal N}_0$ the set of $t$'s for which $p_A(t) = 0$. By continuity of
$p_A(t),~ {\cal N}_0$ is closed and its complement ${\cal N}_0^c$ is
open. Eq. (\ref{t5}) now implies
\begin{equation}\label{t6}
A^{1/2} \psi_t = 0~\mbox{for}~t \in {\cal N}_0~.
\end{equation}
For fixed $\phi \in {\cal H}$ we define the function $F_\phi (z)$ for
Im~$z \le 0$ by
\begin{equation}\label{t7}
F_\phi (z) = \langle \phi, A^{1/2} e^{- i H z} \psi_0 \rangle~.
\end{equation}
By the above remark on $\exp\{-i H z\}$, $F_\phi (z)$ 
is a continuous function for Im~$z \le 0$ and is analytic
for Im~ $z < 0$. By Eq. (\ref{t6}) one has 
\begin{equation}\label{t8}
F_\phi(t) = 0~~~\mbox{for}~ t \in {\cal N}_0~.
\end{equation}

Now let us assume that the complement ${\cal N}_0^c$ is not dense. Then
${\cal N}_0$ contains some interval $I$ of nonzero length, and $F_\phi
(z)$ vanishes on $I$. One can now directly employ the Schwarz
reflexion principle \cite{Levinson} to conclude that $F_\phi (z)
\equiv 0$ or proceed in a more pedestrian way as follows. One defines
an extension of $F_\phi$ to the upper half plane by putting
\begin{equation}\label{t9}
F_\phi (z) = F_\phi(z^*)^* ~\mbox{for}~ \mbox{Im}~z > 0~.
\end{equation}
Since $F_\phi(t) = 0$ for $t \in I$ and thus, a fortiori, real for $t
\in I$, the extension $F_\phi(z)$ continuous for $z \in I$. From this
one easily shows \cite{Levinson} that $F_\phi(z)$ is analytic for $z
\notin I\!\!R \backslash I$, and thus $I$ is contained in the domain
of analyticity. Since $F_\phi(z)$ vanishes on $I$ it must therefore
vanish on the analyticity domain, i.e. for $z \notin I\!\!R \backslash
I$. By continuity $F_\phi(z)$ then vanishes everywhere. Since $\phi$
was arbitrary, we obtain $A^{1/2} \psi_z = 0$ for all $t$. Hence,
\begin{equation}\label{t10}
A \psi_t = 0~~{\rm for~all}~~t
\end{equation}
and thus $p_A(t) \equiv 0$ if ${\cal N}_0^c$ is not dense,
i.e. alternative (ii) holds in this case.

Since a dense open set need not have full Lebesgue measure, it remains
to show that ${\cal N}_0$ is a null set if alternative (ii) does not
hold. To prove this we use the fact that, as a boundary value of a
bounded analytic function, $F_\phi(t)$ satisfies the inequality
\cite{Garnett} 
\begin{equation}\label{t11}
\int_{-\infty}^\infty dt ~\frac{{\rm ln} | F_\phi(t)}{1 + t^2}~> - \infty
\end{equation}
unless it vanishes identically. If ${\cal N}_0$ had positive measure
the integral would be $- \infty$, and thus $F_\phi(t)$ would vanish for all $t$,
for each $\phi$. This would again imply alternative (ii). This proves
the theorem.\\[.5cm]

The theorem is a more abstract version of a result in \cite{He65} on
Fermi's two-atom problem. To check the speed of light in quantum
electrodynamics, Fermi had considered two atoms, separated by a
distance $R$ and with no photons present initially. One of the atoms
was assumed to be in its ground state, the other in an excited
state. The latter could then decay with the emission of a
photon. Fermi calculated the excitation probability of the atom which
had initially been in its ground state. Using standard approximations he
found the excitation probability to be zero for $t < R/c$.

Now, if one takes for $\psi_0$ in the theorem the initial state
considered by Fermi and for $A$ the operator describing the
excitation probability, e.g. the projector onto the excited states,
then $p_A(t)$ becomes the excitation probability, and the theorem
states that this probability is immediately nonzero. Already in
\cite{He65} it was discussed how to avoid a possible conflict with
causality, and this was continued in more detail for example in
\cite{He66,BY,Passante,Milonni}. The upshot was that the immediate
excitation could be understood in a field-theoretic context through
vacuum fluctuations due to virtual photons. The part of the excitation
due to the second atom behaves causally
\cite{Passante,Milonni}. Causality then holds for expectation values
after the spontaneous part has been subtracted. This corresponds to
the notion of weak causality, i.e. for expectation values, introduced in
\cite{II6}, which contrasts to the notion of strong causality, i.e. 
causality for
individual events, as discussed in \cite{He66}. Fermi seems to have had
strong causality in mind.

\vspace*{1.3cm}

\noindent {\bf 3. Application to particle localization and spreading}
\vspace*{.7cm}

We will now apply the above theorem to the question of particle
localization. We note that the results hold independent of whether the
theory is relativistic or not, or a field theory or not. Neither is the
existence of a position operator assumed. The ingredient is just
positivity of the energy. Translation invariance, which was used in
previous treatments \cite{II7,II10,II11,He55}, is not needed here.

Let us suppose that it makes sense to speak of particles inside some
volume $V$, i.e. of the probability to find a particle at a given time
in $V$. This is a highly nontrivial assumption. Indeed, it has been
argued \cite{Redhead} that in algebraic quantum field theory the
notion of local particle states may make sense only asymptotically for
free particles.

In a quantum theory the probability to find a particle or system
inside $V$ should be given by the expectation of an operator, $N(V)$
say. Since probability lie between $0$ and $1$, one must have
\begin{equation}\label{V1}
0 \leq N(V) \le 1~.
\end{equation}

Now let us assume that the system, with state $\psi_0$ at $t = 0$, is
strictly localized in a region $V_0$, i.e. with probability $1$, so
that
\begin{equation}\label{V1a}
\langle \psi_0, N(V_0) \psi_0 \rangle = 1
\end{equation}
or, equivalently,
\begin{equation}\label{V2}
\langle \psi_0, (1 - N(V_0)) \psi_0 \rangle = 0~.
\end{equation}
From Eq. (\ref{V1}) one has
\begin{equation}\label{V3}
1 - N(V_0) \ge 0
\end{equation}
and hence the theorem can be applied, with $A \equiv 1 -
N(V_0)$. As a consequence one either has
\begin{equation}\label{V4}
\langle \psi_t, N(V_0) \psi_t \rangle \equiv 1~~~{\rm for~all}~~ t
\end{equation}
or
\begin{equation}\label{V5}
\langle \psi_t , N(V_0) \psi_t \rangle < 1~~~{\rm for~almost~all}~~t~.
\end{equation}

The argument in Eqs. (\ref{V2}-\ref{V5}) is for pure states. It can easily
be carried over to mixtures characterized by density
matrices. Eq. (\ref{V1a}) is then replaced by 
\begin{equation}\label{V6}
tr (\rho_0 N(V_0)) = 1~.
\end{equation}
Writing
\begin{equation}\label{V7}
\rho_0 = \sum~\alpha_i |\psi_{i 0}\rangle \langle \psi_{i 0}|
\end{equation}
with $\sum \alpha_i = 1$, Eq. (\ref{V6}) becomes 
\[
\sum_i~\alpha_i \langle \psi_{i 0}, N(V_0) \psi_{i 0} \rangle = 1
\]
which implies
\[
\langle \psi_{i 0}, N(V_0) \psi_{i 0} \rangle = 1~~{\rm for~all}~i~.
\]
Then one can proceed as before and obtains that either
\begin{equation}\label{V8}
tr (\rho_t N(V_0)) \equiv 1~~~{\rm for~all}~t
\end{equation}
\begin{equation}\label{V9}
tr(\rho_t N(V_0)) < 1 ~~~{\rm for~almost~all}~t~.
\end{equation}
Alternative (\ref{V8}) or (\ref{V4}) means that the particle or system
stays in $V_0$ for all times, as might happen for a bound state in an
external potential.

Now, if the particle or system is strictly localized in $V_0$ at $t =
0$ it is, {\em a fortiori}, also strictly localized in any larger
region $V$ containing $V_0$. If the boundaries of $V$ and $V_0$ have a
finite distance and {\em if finite propagation speed holds} then the
probability to find the system in $V$ must also be $1$ for
sufficiently small times, e.g.  $0 \le t < \epsilon$. But then the
theorem, with $A \equiv 1 - N(V)$, states that the system stays
in $V$ for {\em all} times. Now, we can make $V$ smaller and let it
approach $V_0$. Thus we conclude that if a particle or system is at
time $t = 0$ strictly localized in a region $V_0$, then finite
propagation speed implies that it stays in $V_0$ for all times and
therefore prohibits motion to infinity. Or put conversely, if there
exist particle states which are strictly localized in some finite
region at $t = 0$ and later move towards infinity, then finite
propagation speed cannot hold for localization of particles.

This can be formulated somewhat more strongly as follows. If at $t =
0$ a particle is strictly localized in a bounded region $V_0$ then,
unless it remains in $V_0$ for all times, it cannot be
strictly localized in a bounded region $V$, however large, for any 
finite time interval thereafter, and the particle localization immediately
develops infinite "`tails"'. The spreading is over all space except
possibly for "`holes"' which, if any, will persist for all times, by
the same arguments as before. If the theory is translation invariant
then there can be no holes, as shown in \cite{II10} under some mild
spectrum conditions. 

\vspace*{1.3cm}

\noindent{\bf 4. Discussion. Connection with causality}

\vspace*{0.7cm}

As shown above, a particle or system, if initially strictly localized
in a bounded region, will immediately develop infinite tails except in
the exceptional case that it stays in the original region for all
times. The latter
seems to require external potentials, and this will be disregarded
here. If a particle is part of a larger system, e.g. of a composite
system, the results still apply, and the appearance of immediate
infinite tails may be attributed to the center-of-mass motion.

In nonrelativistic quantum mechanics the immediate spreading of wave
functions over all space 
is a well known phenomenon. In a relativistic theory this
might lead to a conflict with Einstein causality if it were
observable. Indeed, if a particle were initially strictly localized in
some region on earth and if there were a nonzero probability, however
small, to observe the particle a fraction of a second later on the
moon, this could be used for the absolute synchronization of
clocks. One simply would have repeat the experiment sufficiently
often, preferably with distinguishable particles in order to be sure
when a detected particle was originally released. 

But isn't the Dirac equation for a free particle of spin 1/2 a
counterexample to our results? This is, however, not so. Indeed, the
Dirac equation  is hyperbolic and thus satisfies finite propagation
speed. If for the localization operator $N(V)$ one takes the
characteristic function $\chi_V({\bf x})$ then, for a wave function
with initial support in a finite region, the localization does evolve
causally. However, the Dirac equation contains positive and negative
energy states, and we therefore can conclude from our results that
positive-energy solutions of the Dirac equation always have infinite
support (cf. also \cite{Grosse}).

This example suggests a simple solution to the causality problem
seemingly connected with particle localization. If there would not
exist any 
particle states localized with probability 1 in a bounded region or,
more generally, if all systems were spread out over all space to begin
with, then no problems would arise. This would imply in particular that $N(V)$
could not have the eigenvalue 1 and thus could not be a projector. As a
consequence there would be no selfadjoint position satisfying causal
requirements, not even if one allowed position operators with
noncommuting components as suggested, e.g., in \cite{Bacry}. For if
one had a selfadjoint component of a position operator then its
spectral decomposition would yield localization operators $N(V)$ for
$V$'s being infinite slabs, and to these the results could be applied
in the same way as to bounded regions. 

If one adopts the standpoint that all particle states have infinite
tails to begin with, one might also argue that these tails could be
made to drop off as fast as one likes, thus approximating a strictly
localized state to arbitrary accuracy. In \cite{II11} it was 
shown, however, that in a relativistic theory such tails cannot drop
off arbitrarily fast.

In a field theoretic context the permanent infinite tails could be
intuitively understood through clouds of virtual particles around real
particles ("`dressed particle states"'). Sometimes it is simply argued
that local commutation or anti-commutation relations of the fields
must clearly ensure causal behavior of localized particles. This
overlooks the possibility that the operator $N(V)$ -- if it exists --
might not be a local function of the fields. A more satisfactory discussion
of causality aspects is given in algebraic quantum field theory in \cite{II6}
and in an alternative algebraic framework in \cite{Neumann}.

Instead of speaking about infinite tails one may also envisage that
all particle detectors exhibit inherent noise due to vacuum
fluctuations and that therefore localization with probability 1 or
zero can never be recorded. This would essentially lead to the same
conclusion as permanent infinite tails.

Finally, it may well be true that, as advocated in \cite{Redhead}, the
notion of localizable particles in field theory makes sense only for
free particles.

\end{document}